\documentclass[prl,twocolumn,showpacs]{revtex4}
\advance\textheight by 0.14in
\advance\topmargin by -0.1in

\def\q{{\bf q}}

\def\r{{\bf r}}

\def\u{{\bf u}}

\def\K{{\bf K}}
\def\ltsim{\vbox {\hbox{\lower .8\baselineskip \hbox{$<$}} \break
                 \hbox{\lower 0.2\baselineskip \hbox{$\sim$}} } }
\def\gtsim{\vbox {\hbox{\lower .8\baselineskip \hbox{$>$}} \break
                 \hbox{\lower 0.2\baselineskip \hbox{$\sim$}} } }
\begin{document}

\title{Quantum Depinning Transition of Quantum Hall Stripes}
\author{M.-R. Li$^{1,2,3}$, H.A. Fertig$^{2}$, R. C\^{o}t\'{e}$^{1}$, 
Hangmo Yi$^{4}$}
\affiliation{
$^{1}$D\'epartement de Physique, Universit\'e de Sherbrooke, Sherbrooke, 
Qu\'ebec, Canada J1K 2R1\\  
$^{2}$Department of Physics and Astronomy, University of Kentucky, 
Lexington, Kentucky 40506-0055\\ 
 $^{3}$Department of Physics, University of Guelph, Guelph, 
Ontario, Canada N1G 2W1\\
$^{4}$Korea Institute for Advanced Study, 207-43 Cheongnyangni 2-dong,
Seoul 130-722, Korea}
\date{\today}
\begin{abstract}
We examine the effect of disorder on the electromagnetic response of quantum Hall 
stripes using an effective elastic theory to describe their low-energy dynamics, 
and replicas and the Gaussian variational method to handle disorder effects. Within 
our model we demonstrate the existence of a depinning transition at a critical 
partial Landau level filling factor $\Delta\nu_c$. For $\Delta\nu<\Delta\nu_c$, 
the pinned state is realized in a replica symmetry breaking (RSB) solution, and 
the frequency-dependent conductivities in both perpendicular and parallel to the 
stripes show resonant peaks. These peaks shift to zero frequency as 
$\Delta\nu\rightarrow \Delta\nu_c$. For $\Delta\nu\ge\Delta\nu_c$, we find a 
{\em partial RSB (PRSB)} solution in which there is free sliding only along the 
stripe direction. The transition is analogous to the Kosterlitz-Thouless phase
transition.

\end{abstract}
\maketitle

There is strong evidence from DC transport experiments \cite{stripesexpr} that 
states with stripe ordering form at certain fillings of higher Landau levels 
($N\ge 2$) \cite{stripestheory}. These states exhibit a highly anisotropic, and 
apparently metallic, conductivity \cite{stripesexpr}.  At low temperatures, as 
the partial filling of the highest occupied Landau level, $\Delta \nu$, moves 
away from $1/2$, the electrons in this level cease to contribute to the transport 
properties, and the system behaves in a way typical of the integer quantum Hall 
effect \cite{iqhe}.  One likely interpretation of this change in behavior is that 
the electrons in the partially filled Landau level become pinned by disorder when 
$\Delta\nu < \Delta\nu_c$. The nature of this transition is the subject of this 
work.

Microwave absorption measurements \cite{muwave} provide additional information 
about these systems. These experiments probe the dynamical conductivity of the 
system, $\sigma_{\alpha\beta}(\omega)$, which, in pinned systems, exhibits a peak 
at a frequency determined by the effective restoring force due to the disorder 
\cite{FL1}. Existing data \cite{muwave,Florida} are suggestive of such a peak 
moving to zero frequency as the transition is approached from below, consistent 
with qualitative expectations for a quantum depinning transition \cite{YFC}. 
Interesting filling factor dependences of this peak have also been observed 
in Landau levels where a Wigner crystal is presumably pinned \cite{chen}.

To examine the possibility of a depinning transition in the stripe state, we 
calculate the frequency-dependent conductivity using the replica trick and the 
Gaussian variational method (GVM), first introduced by M\'{e}zard and Parisi 
\cite{MP} and further developed by Giamarchi, Le Doussal, and their coworkers 
\cite{GLD96,GO,CGLD,OC}. In the replica approach, a pinned state is represented 
by one in which there is replica symmetry breaking (RSB). We demonstrate that 
for the stripe system, this leads to peaks both in Re$\sigma_{xx}(\omega)$ and 
in Re$\sigma_{yy}(\omega)$, albeit at slightly different frequencies and with 
different line shapes.  (Here and in what follows, we adopt coordinates such 
that the stripes lie along the $\hat{y}$ direction.) An example of our results 
is illustrated in Fig.~1. A prominent feature of the result is that the peak 
positions move to zero frequency as $\Delta\nu \rightarrow \Delta\nu_c$ from below.

For $\Delta\nu > \Delta\nu_c$, we find a different type of state in which the 
system is pinned for motion perpendicular to the stripes, but is free to slide 
along them.  We call this a {\it partial replica symmetry breaking} (PRSB) state.  
The PRSB state has a number of striking properties, including a power law 
dependence of Re$\sigma_{xx}(\omega) \sim \omega^{\gamma}$ as a function of 
frequency with $\gamma$ continuously increasing with $\Delta\nu-\Delta\nu_c$; 
and a superconducting response \cite{fradkin} at zero frequency in 
Re$\sigma_{yy}(\omega)$, followed by an incoherent metallic response at finite 
frequency. The transition is in the Kosterlitz-Thouless universality class 
\cite{YFC,fertig99}. It exhibits a jump in the low-frequency 
exponent $\gamma$ at the transition, analogous of the universal jump in the 
stiffness of a thin-film superfluid and in the critical exponent of correlation 
functions \cite{nelson}. The possibility of observing these properties in quantum
Hall stripes and other analogous systems is discussed below.

{\em Model and method.}
We start with an action for an elastic system in a magnetic field \cite{YFC} to 
describe the low-energy degrees of freedom of the quantum Hall stripes and their 
nonlinear coupling to the disorder,
\begin{eqnarray}
&& S=S_0+S_{\rm imp}  \label{action} \\
&& S_0= {1\over 2T}\sum_{\q,\omega_n} \sum_{\alpha,\beta=x,y} 
u_\alpha(\q,\omega_n)\, G^{(0)-1}_{\alpha\beta}(\q,i\omega_n) \nonumber \\
&& \;\;\;\;\;\;\;\; \times u_\beta(-\q,-\omega_n)   \label{originalS0} \\
&& S_{\rm imp} = \int d\r \, \int^{1/T}_0 d\tau \, V(\r) \, n(\r,\tau), 
\label{originalSimp} 
\end{eqnarray}
where $u_\alpha$ is a displacement degree of freedom,
$G^{(0)-1}_{\alpha\beta}(\q,i\omega_n)=D_{\alpha\beta}(\q)-
\epsilon_{\alpha\beta}\omega_n/l_B^2$ is the inverse Green's function of $u$
in the pure limit ($\epsilon_{xy} = -\epsilon_{yx} = 1, \epsilon_{xx} = 
\epsilon_{yy} = 0$), $l_B$ is the magnetic length, and $D_{\alpha\beta}(\q)$ is the 
dynamical matrix. This last quantity is determined \cite{YFC} by matching the electron 
density-density correlation function obtained from the elastic model with that 
computed from microscopic time-dependent Hartree-Fock (HF) calculations \cite{CF}. 
In the low-energy sector, $D_{\alpha\beta}(\q)$ has a smectic form, 
$D_{xx}(\q)\simeq d_{xx}(q_x)+\kappa_b q_y^4$, $D_{xy}(\q)=
D_{yx}(\q)\simeq d_{xy}(q_x) q_y$, and $D_{yy}(\q)\simeq d_{yy}(q_x)q^2_y$. 
We note that alternative estimates of $D$ were made by using an edge
state model for the stripe system \cite{edgestatemodel}, which leads to different
results than ours. We will comment on this difference below. 
The disorder potential $V(\r)$ in Eq.~(\ref{originalSimp}) is assumed to be 
Gaussian distributed with zero average, $\overline{V(\r)V(\r')}= V^2_0a_xa_y
\delta(\r-\r')$, where $a_x$ and $a_y$ are the lattice constants of the stripe 
crystal. In Eq.~(\ref{originalSimp}) $n(\r,\tau)$ is the electron density operator 
whose Fourier transform we approximate by \cite{GLD96}
$n(\q,\tau) \simeq n_0\big\{1 -i\q\cdot \u(\q,\tau) +  
\sum_{\K\neq 0} \int d\r \,e^{i\K \cdot [\r-u(\r,\tau)]-i\q\cdot \r}\big\}$ 
where $n_0$ is the average electron density and $\K$ is a stripe crystal 
reciprocal lattice vector. For simplicity, 
we drop the $i\q \cdot \u$ term in $n(\q,\tau)$  since this cannot
pin the electron system \cite{GLD96}, and 
we keep only the lowest harmonics of the reciprocal lattice vectors, 
$K_x=0, \pm 2\pi/a_x$ and $K_y=0, \pm 2\pi/a_y$. These approximations should not 
qualitatively change our results. 

We use replicas and the GVM to handle the disorder effects. This method has been 
employed with great success in a number of systems \cite{GLD96,GO}, although some 
controversy has arisen over previous applications to quantum Hall systems, for 
reasons that do not apply to our present study \cite{OC,footnote1}. We introduce 
$n$ replicas of the system where eventually we will take $n\rightarrow 0$. This 
involves setting $S_0\rightarrow S_0^{\rm (eff)}=\sum^n_{a=1} S_0^{(a)}$ with 
$a$ the replica index; averaging over disorder introduces coupling among the 
replicas in the form $S^{\rm (eff)}_{\rm imp} \simeq - v_{\rm imp}\sum^n_{a,b=1} \, 
\int^{1/T}_0 d\tau_1 \, d\tau_2 \int d\r \sum_{\K\neq 0} 
\cos\{\K\cdot[\u^a(\r,\tau_1)-\u^b(\r,\tau_2)]\}$, with $v_{\rm imp}=V_0^2 
a^2_x a^2_y n^2_0$. The GVM \cite{MP} consists of replacing $S_0^{\rm (eff)}+
S^{\rm (eff)}_{\rm imp}$ with a variational Gaussian action, characterized by 
variational Green's function $G^{ab}_{\alpha\beta}(\q,i\omega_n)$ which  
is chosen to minimize the free energy. This leads to a set of saddle point 
equations (SPE's).  It is convenient to parameterize the matrix $G$ in terms of 
a self-energy matrix $\zeta$, such that $(G^{-1})^{ab}_{\alpha\beta}(\q,i\omega_n) 
= G^{(0)-1}_{\alpha\beta}(\q,i\omega_n) \, \delta_{ab} - 
\zeta^{ab}_{\alpha\beta}(i\omega_n)$. Note that we can safely assume $\zeta$ has 
no ${\bf q}$ dependence because none emerges in the SPE's. We also set 
$\zeta^{ab}_{xy}=\zeta^{ab}_{yx}=0$ since it is a valid solution of the SPE's, 
and preserves reflection symmetry.

Our SPE's are a natural generalization of those found using the replica and 
GVM on isotropic systems \cite{GLD96,GO,CGLD}. 
Here, we will present only the final equations, deferring details to a future 
publication \cite{LFCY}. After taking the limit $n \rightarrow 0$, the
self-energy is characterized by a replica diagonal component, 
$\zeta^{aa}_{\alpha\alpha}(i\omega_n)\rightarrow \tilde{\zeta}_{\alpha}(i\omega_n)$
and an off-diagonal function, $\zeta^{a \ne b}_{\alpha\alpha} (i\omega_n)
\rightarrow \zeta_{\alpha}(u)\delta_{\omega_n,0}$, where $0<u<1$.  It is the first of these that 
directly enters the frequency-dependent conductivity, while the second of these 
determines whether the system is in a replica symmetric state (constant $\zeta(u)$), 
a RSB state ($\zeta(u)$ varying with $u$) or some other state. The dynamical 
conductivity is obtained \cite{GLD96} via 
\begin{eqnarray}
\sigma_{\alpha\beta}(\omega) =  (e^2/a_x a_y)\, i\omega
\, \widetilde{G}^{\rm ret}_{\alpha\beta}(\q=0,\omega), \label{conductivity}
\end{eqnarray}
with $\widetilde{G}^{\rm ret}_{\alpha\beta}(\q,\omega)=
\widetilde{G}_{\alpha\beta}(\q,i\omega_n \rightarrow \omega + i\delta)$,
where $\widetilde{G}_{\alpha\beta}$ is the $n \rightarrow 0$ limit of 
$G^{aa}_{\alpha\beta}$. The self energy $\tilde{\zeta}^{\rm ret}_{\alpha}(\omega)$ 
is the analytic continuation of $\tilde{\zeta}_{\alpha}(i\omega_n)$, which 
satisfies the following SPE's \cite{LFCY}: 
\begin{eqnarray}
&& \tilde{\zeta}^{\rm ret}_{\alpha}(\omega) 
= e_{\alpha} - \big[F_{\alpha}(\omega)-F_{\alpha}(0^+)\big], \label{SPEw}  \\   
&& F_{\alpha}(\omega)= 4v_{\rm imp} \sum_{\K\neq 0} K_\alpha^2 \int^\infty_0 dt \; 
e^{i\omega t} \nonumber \\
&& \times  {\rm Im} \; {\rm exp} \bigg\{- \sum_{\mu=x,y} 
{K_\mu^2\over \pi} \int^\infty_0 df \, A_{\mu}(f) \big[ 1-e^{itf}\big]
\bigg\}, \label{Fw}  
\end{eqnarray}
where $e_{\alpha}=\int^1_0 du \, \zeta_{\alpha}(u)- F_{\alpha}(0^+)
-2 v_{\rm imp} \sum_{\K\neq 0} K_\alpha^2 \int^{1/T}_0 d\tau 
\exp \{-T\sum_{\mu} K_\mu^2\sum_{\q,\omega_n} 
[1-\cos\omega_n \tau] \, \widetilde{G}_{\mu\mu}(\q,\omega_n)\}$ 
and $A_{\mu}(f)=\sum_\q {\rm Im} \widetilde{G}^{\rm ret}_{\mu\mu}(\q,f)$.

The constants $e_x,~e_y$ may be regarded as a measure of the strength of pinning 
by the disorder potential. When one or both of these vanish, the effective elastic 
constants $G^{-1}({\bf q}=0,\omega_n)$ allow the system to be shifted as a whole 
without any energy cost.  If one knows these constants then Eqs.~(\ref{SPEw}) and 
(\ref{Fw}) form a closed set of equations which may be solved numerically. A common 
further ``semiclassical'' approximation \cite{GLD96} involves expanding 
Eq.~(\ref{Fw}) for small $A_\mu$, which is valid when the fluctuations of the 
stripes are small. While it greatly simplifies the numerics, this approximation 
is invalid near the depinning transition since the fluctuations become arbitrarily 
large. We thus leave Eq.~(\ref{Fw}) in its present form.

We are left with the task of computing $e_x$ and $e_y$. As has been discussed 
elsewhere \cite{GLD96}, this can be accomplished without fully solving the SPE's 
by imposing the condition Re $\sigma \sim \omega^2$ for small $\omega$, which 
guarantees that the collective mode density of state vanishes at zero frequency.
(This constraint can also be justified by requiring marginal stability of the replicon 
mode \cite{GLD96}.)  The second constraint can be found from the SPE's for 
$\zeta_{\alpha}(u)$ with the assumption of a {\em one-step} RSB, which is a common 
solution in low-dimensional systems \cite{GLD96,GO,CGLD}. This leads after some work 
\cite{LFCY} to the condition 
\begin{eqnarray}
e_y/e_x = \sum_{\K\neq 0} K^2_y \, e^{-W(\K)}
/\big[\sum_{\K\neq 0} K^2_x \, e^{-W(\K)}\big], \label{constraint2}
\end{eqnarray}
where $W(\K) = {1\over \pi} \sum_\mu K^2_\mu  \int^\infty_0 df \, A_{\mu}(f)$ are 
Debye-Waller factors. These play a prominent role in the depinning transition: as 
we shall see, $W({\bf K})$ diverges whenever ${\bf K}$ has a component along the 
$\hat{y}$ axis as $\Delta\nu_c$ is approached from below, leading to a suppression 
of $e_y$. This behavior cannot be captured by the semiclassical approximation.

\begin{figure}[h]
\begin{picture}(250,340)
\leavevmode\centering\includegraphics{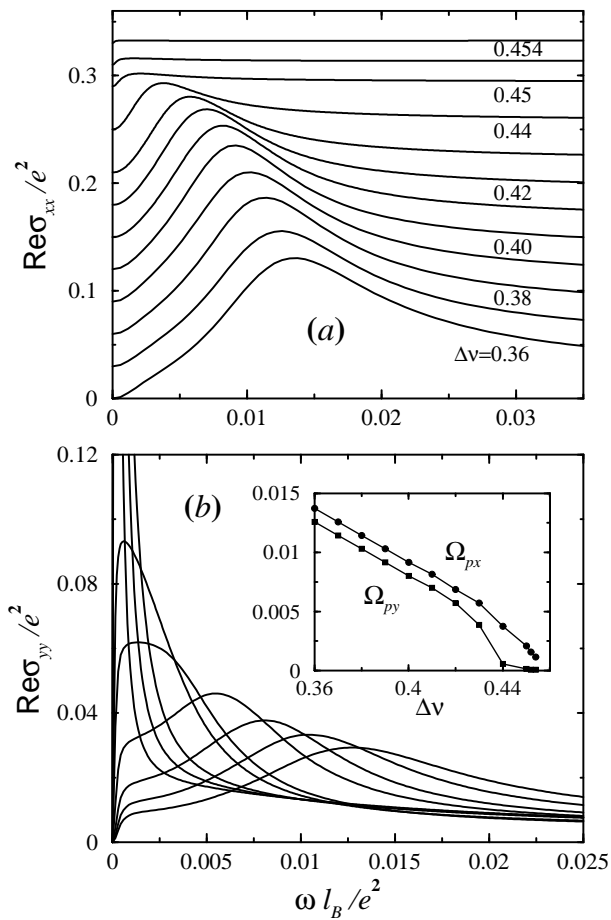}
\end{picture}
\caption{Real parts of conductivities as functions of frequency in the pinned state 
(a) perpendicular to the stripes, (b) along the stripes. The $\hbar=1$ unit 
and $v_{\rm imp}=0.0005e^4/l_B^2$ are used. In (a) all curves start from 
Re$\sigma_{xx}=0$ at $\omega=0$, and curves except for $\Delta\nu=0.36$ are lifted 
upward for clarity.  Curves from right to left in (b) correspond to 
$\Delta\nu=0.36, 0.38, 0.4, 0.42, 0.43, 0.44, 0.45, 0.452, 0.454$, respectively. 
Inset in (b): peak positions $\Omega_{px}$ and $\Omega_{py}$, 
in units of $e^2/l_B$, as functions of $\Delta\nu$.  } 
\label{sgmpinned}
\end{figure}

{\em Results -- RSB solution.} 
We have solved numerically Eqs.~(\ref{SPEw}) and (\ref{Fw}) along with the
two constraints using an iterative method \cite{LFCY}. In what follows we present 
some results for electrons in the $N=3$ Landau level, with 
a disorder level $v_{\rm imp}=0.0005e^4/l^2_B$. This is likely to be 
somewhat larger than experimental values, but we choose it for numerical
convenience and do not expect our results to qualitatively change 
with smaller disorder strengths. The dynamical conductivities for these parameters 
when the system is in a pinned (RSB) phase are presented in Fig.~1.  For $\Delta\nu$ 
well below $\Delta\nu_c \approx 0.46$, Re$\sigma_{xx}$ has a pinning peak
whose lineshape is qualitatively similar to what is found using the semiclassical 
approximation \cite{CGLD}. The prominent behavior visible in Fig.~1~(a) is the 
collapse of the peak frequency $\Omega_{px}\rightarrow 0$ as the depinning 
transition is approached. 
Experimental observations are so far consistent with this\cite{muwave,Florida}.
Re$\sigma_{yy}$ also has a collapsing peak, but the 
observed lineshape is more interesting.  Below the peak frequency $\Omega_{py}$, in 
the range $e_y < \omega < \Omega_{py}$ the conductivity appears to tend toward a 
non-vanishing value when $\Delta\nu$ is sufficiently below $\Delta\nu_c$. The 
quantity $e_y$ turns out to be rather small due to a large Debye-Waller factor, 
and in this frequency range the system displays a behavior similar to the incoherent 
metal response at non-vanishing frequency of the depinned (PRSB) phase which we 
discuss below.  For $\omega \ll e_y$, Re$\sigma_{yy}(\omega)$ vanishes quadratically 
with $\omega$ (not visible on the scale of Fig.~1), as required for a pinned state.  
As $\Delta\nu \rightarrow \Delta\nu_c$, we eventually reach a situation in which 
$e_y$ and $\Omega_{py}$ are of similar order, in which case the pinning peak 
sharpens and grows quite large. This peak continuously evolves into a 
$\delta$-function at zero frequency as the system enters into the PRSB state,
so that the transition from pinned to depinned behavior is very continuous.

{\em Results -- PRSB solution}.
For $\Delta\nu\ge \Delta\nu_c$ the state is characterized by $e_x\ne 0$ but 
$e_y=0$. This corresponds to a RSB solution for  $\zeta_x(u)$ but a replica 
symmetric solution for $\zeta_y(u)$. We call this the {\em partial RSB} (PRSB) state. 
In this situation, the system is pinned perpendicular to the stripe direction, 
but is free to slide along it.  This is consistent with the results from a 
perturbative renormalization group study of the same model \cite{YFC}, where 
the coupling of the disorder to the motion of the system parallel to the stripes 
was shown to be irrelevant when $\Delta\nu$ is close enough to 1/2. This irrelevance
suggests that the PRSB phase should be in a superconducting state. This observation 
is born out by the presence of a $\delta$-function in Re$\sigma_{yy}$ at zero 
frequency. Remarkably, Re$\sigma_{xx}$ vanishes at zero frequency, so that
we find the PRSB state is one with an {\it infinite} anisotropy in the DC 
conductivity. This is not observed in DC transport experiments \cite{stripesexpr}, 
and we comment below on what is missing from our model that leads to this discrepancy.
The possibility of such behavior for quantum Hall stripes was first suggested
in Ref.~[\onlinecite{fradkin}].

The origin of the PRSB and its structure may be understood from the SPE 
(\ref{SPEw}). The state is characterized by $e_x\ne 0$ and 
Im$\tilde{\zeta}^{\rm ret}_x(\omega)\sim \omega$ at small $\omega$, and 
$\tilde{\zeta}^{\rm ret}_y(\omega)$ vanishing faster than linearly in $\omega$. 
It is easy to show in this situation, $A_y(f)\sim 1/f$ at small $f$. After some 
algebra, we find \cite{LFCY} that when  
\begin{eqnarray}
&& \gamma = {a_yl_B\over a_x} \int d{q_x} {d_{xx}-
e_x\over \sqrt{[d_{xx}-e_x]d_{yy}-d^2_{xy}}}-2
\ge 1,   \label{PRSBexponent}
\end{eqnarray}
a self-consistent solution of Eq.~(\ref{SPEw}) emerges, with 
${\rm Re~}\tilde{\zeta}^{\rm ret}_y(\omega) \sim \omega^2$ and 
${\rm Im~}\tilde{\zeta}^{\rm ret}_{y}(\omega) \sim \omega^{\gamma+1}$.  
The large value of this last exponent leads both to
a power law dependence of Re$\sigma_{xx}\sim \omega^{\gamma}$ and the 
$\delta$-function response at zero frequency in  Re$\sigma_{yy}$.
It is also easy to show that $\lim_{\omega \rightarrow 0}
{\rm Re}\sigma_{yy}$ = const.$>0$, so that, at low but finite
frequency the response is {\it metallic}.  This combination
of infinite and ``incoherent'' metallic response is typical
of superconductors \cite{kohn}.  

In the vanishing disorder limit, the minimum value of $\gamma$ from the formula 
(\ref{PRSBexponent}) for which we can obtain a PRSB solution agrees precisely with
the condition found in Ref.~\onlinecite{YFC} for pinning along the stripes to become 
irrelevant. One remarkable consequence of this limiting value is that 
Re$\sigma_{xx}\sim \omega^{\gamma}$ with $\gamma \rightarrow 1$ as the transition 
is approached from above, whereas just below the transition we expect, in the 
pinned state, Re$\sigma_{xx}\sim \omega^{2}$. Thus, the low frequency exponent 
{\it jumps} at the transition, in a way that is analogous to the universal jump 
in the superfluid stiffness and the critical exponent of correlation functions of 
the Kosterlitz-Thouless transition \cite{nelson}. This behavior is similar to 
what happens in the roughening transition \cite{fertig99,chaikin}.

In real DC transport experiments, one observes a finite anisotropy rather 
than the infinite one found in the PRSB state. The missing ingredients from 
our model are processes allowing hopping of electrons between stripes.
These processes are very difficult to incorporate into an elastic model.  
It is clear that, if relevant, such processes can broaden the 
$\delta$-function response to yield anisotropic metallic behavior.
Our results should apply at frequency scales above this 
broadening. Indeed, microwave absorption experiments 
become quite challenging at low frequencies, and it is unclear
whether existing measurements of the dynamical conductivity 
can access the low frequency conductivity in the unpinned state, whether or 
not it is broadened. In any case, it is interesting to speculate that a true 
$\delta$-function response might be accessible in structured environments 
where barriers between stripes may suppress electron hopping among stripes
\cite{Endo}, or that there may be analogous states for layered 2+1 dimensional 
classical systems of long string-like objects, which has been shown \cite{fertig99} 
to be closely related to the two-dimensional quantum stripe problem.

The quantum depinning transition we find is unlikely to occur in models
which preserve particle-hole symmetry (PHS) at $\Delta\nu=1/2$ \cite{edgestatemodel}.
Our model overcomes this limitation because the HF state we use
spontaneously breaks PHS at this filling to arrive at a lower energy
state than the simpler ``box-filled'' state \cite{stripestheory}
which has been used in the edge state description of the
quantum Hall smectic \cite{edgestatemodel}.
It is at present unclear if quantum fluctuations restore PHS to the 
quantum Hall smectic at $\Delta\nu=1/2$.  Our results
offer a falsifiable experimental test that can settle this question.

{\em Acknowledgements.}
The authors are especially grateful to R.~Lewis, L.~Engel, and Y. Chen for 
many stimulating discussions about this problem, and for showing us their 
experimental data prior to publication. We are also indebted to G.~Murthy, 
E.~Orignac, E.~Poisson, and A.H. MacDonald for useful discussions and 
suggestions.  This work was supported by a NSF Grant No. DMR-0108451, by 
a grant from the Fonds Qu\'eb\'ecois de la recherche sur la nature et les 
technologies and a grant from the Natural Sciences and Engeneering 
Research Council of Canada, and by a grant from SKORE-A program.

\end{document}